\documentclass[%
 reprint,
 amsmath,amssymb,
 aps,
 prd
floatfix,
showkeys
]{revtex4-1}

\usepackage[USenglish]{babel}
\usepackage{slashed}
\usepackage{physics}
\usepackage[normalem]{ulem}
\usepackage{graphicx}
\usepackage{dcolumn}
\usepackage{bm}
\usepackage[usenames]{color}
\usepackage{xcolor}
\usepackage[small]{caption}
\usepackage{float}
\usepackage{hyperref}
\usepackage{cleveref}
\usepackage{verbatim}

\newcommand{\be}{\begin{equation}}
\newcommand{\ee}{\end{equation}}
\newcommand{\bea}{\begin{eqnarray}}
\newcommand{\eea}{\end{eqnarray}}
\newcommand{\beas}{\begin{eqnarray*}}
\newcommand{\eeas}{\end{eqnarray*}}

\newcommand{\apjl}{Astrophysical Journal L}
\newcommand{\aap}{Astronomy and Astrophysics}

\newcommand{\sg}[1]{\textsf{\color{black}{ #1}}}

\begin{document}

\title{Tidal heating in binary inspiral of strange quark stars}

\author{Suprovo Ghosh$^{1,2}$, Jos\'e Luis Hern\'andez$^{3,4,5}$, Bikram Keshari Pradhan$^{1,6}$, Cristina Manuel$^{3,4}$, Debarati Chatterjee$^{1}$, and Laura Tolos$^{3,4}$  }
\affiliation{%
$^1$Inter University Centre for Astronomy and Astrophysics, Ganeshkind, Pune 411007, India\\
$^2$Mathematical Sciences and STAG Research Centre, University of Southampton, Southampton SO17 1BJ, United Kingdom\\
$^3$Institute of Space Sciences (ICE, CSIC), Campus UAB,  Carrer de Can Magrans, 08193 Barcelona, Spain\\
$^4$Institut d'Estudis Espacials de Catalunya (IEEC), 08034 Barcelona, Spain \\
$^5$Facultat de Física, Universitat de Barcelona, Martí i Franquès 1, 08028 Barcelona, Spain.\\
$^6$IP2I Lyon, University Claude Bernard Lyon 1, CNRS/IN2P3, UMR 5822,
 69622 Villeurbanne, France
}

\begin{abstract}
We investigate tidal heating associated with the binary inspiral of strange quark stars and its impact on the resulting gravitational wave signal. Tidal heating during the merger of neutron stars composed of nuclear matter may be considered negligible, but it has been demonstrated recently that the presence of hyperons at high densities could  significantly enhance the dissipation during inspiral. In this work, we evaluate the bulk viscosity arising from non-leptonic weak processes involving quarks and show that it can be several orders of magnitude higher than the viscosity of nuclear matter at temperatures relevant to the inspiral phase of the merger of strange stars. We model strange quark matter in the normal phase using a non-ideal bag model including electrons and ensure compatibility with astrophysical constraints. By analysing equal-mass binary systems with component masses ranging from 1.4 to 1.8 $\, M_{\odot}$, we find that temperatures close to 0.1 MeV are reached by the end of the inspiral phase. We also estimate the effect on the gravitational waveform and conclude that the additional phase shift could range from $0.1$ to $0.5$ radians for strange quark masses of 200 MeV, making it potentially detectable by next-generation gravitational wave detectors. Given that tidal heating from hyperons is dominant only for very massive neutron stars having masses 1.8 to 2.0 $\, M_{\odot}$, a successful detection of this phase shift during the inspiral of binary systems with relatively low masses of 1.4 to 1.6 $\, M_{\odot}$ could be a smoking gun signature for the existence of strange quark stars.

\end{abstract}
\maketitle

\section{Introduction}
\label{sec:intro}

Compact stars offer unique astrophysical systems where cold matter at ultra high density can be studied, given the high complexity to reproduce such conditions in terrestrial experiments.  
It is well known that dense matter at the lowest densities in these environments is made of neutrons, protons and electrons, but as the density increases to a few times the nuclear saturation density inside the core of these compact objects, it is challenging to predict its composition. According to Quantum Chromodynamics (QCD), deconfinement to quarks and gluons in the so-called quark-gluon plasma is expected at high densities~\cite{BAYM1976241,Chapline:1976gq,Collins:1974ky,PhysRevD.16.1169}. However, at typical densities of compact stars the theory is still strongly coupled making it difficult to determine the phases of dense matter~\cite{HEISELBERG2000237,WEBER2005193,Alford_2008, Blaschke2018}. 

The latest theoretical, experimental and astrophysical developments in multi-messenger physics provide opportunities to infer the composition and other properties of these objects. The recent observation of the binary neutron star (BNS) merger event GW170817~\cite{Abbott2017,Abbott2018,Abbott2019} by the LIGO–Virgo GW detectors~\citep{AdvLIGO2015,Abbott2016,AdvVIRGO2014} along with electromagnetic (EM) counterparts~\citep{MultiGW170817} have provided significant constraints on the unknown Equation of State (EoS) of dense matter~\cite{Annala,De,Most,Dietrich2020,Pang_2021,Issac2021,Tolos:2020aln,Burgio:2021vgk}. With the increased sensitivity of current GW detectors ~\citep{Aplus_sensitivity} (hereafter referred as A+) or with the proposed next generation GW detectors such as Einstein Telescope (ET)~\cite{Hild_2011} and Cosmic Explorer (CE) ~\citep{CE,CE2}, possible detection of more gravitational wave events in the inspiral as well as the post-merger phase of binary coalescence of neutron stars offers exciting opportunities to put constraints on the equilibrium EoS and as well as its transport properties ~\cite{Kokkotas:1999bd,Rezzolla:2003ua,Sieniawska:2019hmd,Sotani2011,Pradhan2023,Most_2021,Celora_2022,Most_2024,Chabanov:2023blf},
which strongly depend on the constituent particles of the star and their interactions.

The effect of the tidal interaction between the components of a neutron star binary on the emitted gravitational waveform has been crucial for extracting information about the dense matter behaviour and composition from GW data~\cite{Abbott2017,Annala,De,Most,Dietrich2020,Pang_2021,Ghosh_2022,Ghosh2022}.  Adiabatic tides or tidal deformability have the dominant effect on the gravitational waveform towards the late inspiral~\cite{Hinderer}. Dynamical tides that correspond to the resonant or non-resonant excitation of the internal modes of the neutron stars can also become significant depending on the mode frequencies and their coupling to the tidal field~\citep{Andersson_2018,Schmidt2019}. It has been shown that neglecting dynamical tidal corrections can significantly bias the tidal and EoS inference from future BNS events \citep{Prattendynamical,Pradhan2023,Pradhan2022}, detected with the increased sensitivity of current GW detectors, as well as with upcoming next-generation detectors.

These conservative tidal effects (adiabatic and dynamic) probe only the equilibrium EoS. However, these tidal interactions can also drive the system out of equilibrium depending on the relevant nuclear process timescales. These out-of-equilibrium viscous processes inside the star damp out the tidal energy and convert it to heat which we refer to as ``tidal dissipation" or equivalently ``tidal heating". Earlier estimations~\cite{Lai_1994,Bildsten1992,Arras2019} suggested that the viscous damping of the stellar oscillation modes from  viscosity of nuclear matter i.e., shear viscosity from $ee$ scattering and bulk viscosity from Urca reactions~\cite{Sawyer1989} are not significant enough to be considered for GW studies during the inspiral phase of the merger. Although most of the existing studies on tidal dissipation of binary neutron stars have been performed in the Newtonian limit, in a few recent works  ~\citep{Ripley_2023,Ripley_2024,r2024dynamical,r2024dissipativetidaleffectsnexttoleading} the signature of tidal lag in GWs was re-analysed in an effective relativistic theory of tidal response and also in a fully relativistic formalism using gravitational Raman scattering~\citep{Saketh_2024}. 

All these previous works considered only the hadronic sources of shear viscosity originating from $ee$ scattering as the dominant source of shear viscous damping at low temperatures ($T \ll  10^{10}$ K which is $\approx 1$ MeV) during the inspiral. But, as explained earlier, the densities at the core of neutron stars can be high enough for strange matter such as hyperons or deconfined quark matter to appear as stable components. Recently, Ghosh et al.~\cite{Ghosh_2024} demonstrated that the bulk viscosity originating from non-leptonic weak processes involving hyperons can be several orders higher ($\approx 10^{8} - 10^{10}$ times) than shear viscosity from $ee$ scattering in the temperature range of $10^{6}-10^8$ K~\cite{Lindblom2002,Debi_2006,Alford2021}, and the dissipation of the dominant $f-$mode introduces an additional phase of the order $10^{-3} - 0.5$ rad to the GW signal depending on the neutron star masses, which should in principle be detectable by the next generation GW detectors~\citep{ghosh2025tidaldissipationbinaryneutron}.

Based on the fact that hyperonic degrees of freedom result in large viscous dissipation and consequently tidal heating, a dense medium composed of strange quark matter in the normal phase offers a potential scenario where the conversion of tidal energy to heat may also be relevant. Additionally, it is expected that viscous dissipative effects in strange stars, composed entirely of quark matter, would be higher than in  hybrid stars where quarks are only present in the high density core. 

In general, viscous dissipation in strange quark stars is driven by shear and bulk viscosities. \textcolor{black}{Shear viscosity generated by quark-quark scattering due to gluon exchange has been calculated up to $\mathcal{O}(\alpha_s)$ in a standard Fermi liquid approximation~\cite{haensel1989transport} and corrected in \cite{PhysRevD.48.2916}, taking into account that the dynamical screened transverse gluon interactions lead to a non-Fermi liquid behavior.}  
At densities and mode frequencies found in compact stars \textcolor{black}{and at temperatures ranging from $10^7$ to $10^{10}$ K}, shear viscosity results to be orders of magnitude smaller than the bulk viscosity, the latter being more relevant in the tidal energy dissipation. 

Currently extensive literature exists on bulk viscosity in strange quark matter addressing different quark matter phases~\citep{Harris:2024evy}, including the early studies in Refs.~\cite{SAWYER1989412, Madsen:1992sx}. Particularly, strange quark matter in the normal phase has been studied employing bag models~\cite{Alford_2006, Alford_2010, PhysRevD.109.123022}, density-dependent quark mass models~\cite{PhysRevC.70.015803}, using leading corrections due to strong interactions from perturbative QCD~\cite{Sad:2007afd, PhysRevD.109.123022} and also at next-to-next-to leading order~\cite{Cruz_Rojas_2024}. Other works in unpaired strange quark matter consider a finite magnetic field~\cite{Huang_2010} and anharmonic density oscillations~\cite{Shovkovy_2011}.
All the studies mentioned above agree that at low temperatures (below than $10^9$ K) and frequencies around $1$ kHz, the bulk viscosity reaches its maximum receiving the highest contribution from non-leptonic weak processes. Thus, we expect that bulk-viscous dissipation from strange quark stars may introduce enough additional phase correction to the frequency domain gravitational waveforms during the inspiral that can possibly be detectable with the next generation GW detectors.

This article is structured as follows: in Sec.~\ref{sec:TIDALDISSIPATEDENERGY} we briefly recapitulate the  Newtonian tidal heating formalism during the inspiral stage of the coalescence of BNS and estimate the bulk viscous dissipated energy. We also present the non-ideal bag model we use to describe unpaired strange quark matter in subsection~\ref{subsec:EOS_QM}, the
strange star EoS in subsection~\ref{subsec.strangestarEoS}, and calculate the bulk viscosity associated with non-leptonic weak interactions
among quarks in subsection~\ref{subsec:bulkviscosity}, which are used to calculate the tidal dissipated energy in the binary inspiral. In Sec.~\ref{sec:Temp}, we estimate the temperature reached during the inspiral due to this dissipation from strange quark bulk viscosity. Then in Sec.\ref{sec:phase}, we estimate the additional phase correction introduced in the frequency domain gravitational waveform due to this viscous dissipation and compare with the current limits with future generation GW detector sensitivity to argue about the detectability. In Sec.~\ref{sec:discussion} we discuss the main implications of this work and also future directions. \textcolor{black}{We use natural units, assuming $G=\hbar=c =k_B= 1$, unless explicitly stated otherwise}.

\section{Tidal dissipated energy from mode damping}
\label{sec:TIDALDISSIPATEDENERGY}
In this section, we describe the formalism for the Newtonian tidal heating in the BNS inspiral from linear perturbations of a background solution for a star in equilibrium following the Refs.~\cite{Ghosh_2024,Lai_1994}. During the inspiral of the BNS system, due to the tidal interactions, various oscillation modes inside the star can be resonantly or non-resonantly excited depending on their frequencies compared to the orbital frequency~\cite{Andersson_2018}. Since the fluid is viscous, a fraction of this tidal energy will be dissipated as thermal energy and will increase the temperature of the system. The effect of this dissipated energy depends on how large the timescale of the viscous dissipation is compared to the orbital timescale. If the dissipation timescale is much longer than the orbital one, then viscous dissipation will not have much effect on the inspiral of the BNS system~\cite{Lai_1994}. \\

Let us first analyse the timescale for viscous mode damping. The mode damping rate is given by
\begin{equation}\label{eq:damptime2}
    \gamma_{\alpha} = \Dot{E}_{\text{visc},\alpha}/(2E_{\alpha}) ,
\end{equation}
where $\alpha(\equiv \{n,l,m\})$ denotes the normal mode index, $E_{\alpha}$ is the energy of the mode and $\Dot{E}_{\text{visc},\alpha}$ is the energy dissipation rate. The rate of dissipated energy is given in terms of the viscous stress tensor $\sigma_{ij}$~\cite{Lai_1994}
\begin{equation}\label{eq:viscenergy}
    \Dot{E}_{\rm visc} = \int d^3x \,\sigma_{ij}\bm{V}_{i,j} ,
\end{equation}
where $\bm{V}_{i,j}$ denotes the derivative of the $i$th component of the perturbation velocity vector $\bm{V}$ w.r.t. to the $j$th co-ordinate and the viscous stress tensor $\sigma_{ij}$ can be written as~\cite{Landau1987Fluid}
\begin{equation}\label{eq:visctensor}
    \sigma_{ij} = \eta_{SV}\left(\bm{V}_{i,j} + \bm{V}_{i,j} - \frac{2}{3}\delta_{ij}\nabla .\bm{V}\right) + \zeta\delta_{ij}\nabla . \bm{V} ,
\end{equation}
where $\eta_{SV}$ and $\zeta$ are the shear and bulk viscosity coefficients, respectively. \\

In the adiabatic approximation, the effect of the tidal potential due to the companion star is measured in terms of the Lagrangian fluid displacement vector $\bm{\xi}(r,t)$ from its equilibrium position. This displacement can be analysed in terms of the normal modes of the neutron star, 
\begin{equation}\label{eq:normal}
    \bm{\xi} (\bm{r},t) = \sum_{\alpha}\bm{\xi_{\alpha}(r)} a_{\alpha}(t) ,
\end{equation}
where $\xi_{\alpha}(r)$ is the eigenfunction \textcolor{black}{with the orthogonality and normalization condition
\begin{equation}
 \int d^3x \,\rho\, \xi^*_{\alpha}\xi_{\alpha'} = \delta_{\alpha\alpha'}
\end{equation}
where $\rho$ is the energy density}, and $a_{\alpha}(t)$ is the amplitude of the particular eigenmode due to the tidal field of the companion and as stated before, $\alpha$ is the normal mode index.  For a particular mode, the eigenfunction can be written as the sum of the radial $\xi^r_{nl}$ and tangential $\xi^{\perp}_{nl}$ components 
\begin{equation}\label{eq:eigenvector}
    \bm{\xi_{\alpha}}(r) = \left[\xi^r_{nl}(r)\bm{e_r} + r\xi^{\perp}_{nl}(r)\bm{\nabla}\right]Y_{lm}(\theta,\phi) ,
\end{equation}
where $\bm{e_r}$ is the radial vector and $Y_{lm}(\theta,\phi)$ are the spherical harmonic functions. To leading order, when the viscous dissipation rate is smaller than twice the mode angular frequency ($\gamma_{\alpha} < 2\omega_{\alpha}$), using the expression of the displacement vector in Eq.~\eqref{eq:normal} and $ \bm{V} = \frac{d}{dt}\bm{\xi} (r,t)$ in Eq.~\eqref{eq:viscenergy}, we can get the dissipated energy during the inspiral as 
\begin{equation}\label{eq:viscenergy_final}
    \Dot{E}_{\rm visc} \approx \sum_{\alpha}2\gamma_{\alpha}\Dot{a}_{\alpha}(t)^2.
\end{equation}
In this work, we will only consider the dominant $l = 2$ fundamental $(f)$ mode contribution (dropping the subscript $\alpha$ henceforth) to the tidal dissipation because it has the strongest coupling to the external tidal field, although estimates by Lai~\cite{Lai_1994} showed that the leading order gravity $g$ mode dissipation can be also significant. 

Then, we can express the bulk viscous dissipation rate as 
\begin{equation}\label{eq:gammabulk}
    \gamma = \frac{1}{2} \frac{(l+|m|)!}{(l-|m|)!}\int_0^R r^2dr\zeta\left(\frac{\partial \xi^r}{\partial r} + \frac{2}{r}\xi^r - l(l+1)\frac{\xi^{\perp}}{r}\right)^2 ,
\end{equation}
where $\xi^r$ and $\xi^{\perp}$ are the radial and perpendicular components of the $f$-mode eigenfunction, and the corresponding velocity field can be written as  \textcolor{black}{$\bm{V} = -i\omega \bm{\xi}$}, where $\omega$ is the $f$-mode angular frequency, which is determined via a relativistic Cowling approximation~\cite{Pradhan_2021} as outlined in~\cref{app:f_mode}. The normalised mode eigenfunctions are also used to calculate the \textcolor{black}{tidal coupling strength} $Q_{nl}$ defined as 
\begin{equation}
   Q_{nl} = \int_0^R \rho lr^{l+1} [\xi_{nl}^r(r)  + (l+1)\xi_{nl}^{\perp}(r)]dr ,
\end{equation}
where $R$ is the radius of the star. In the adiabatic limit away from the merger when the orbital evolution is sufficiently slow ($\dot{\Omega}/\Omega \ll 1$ where $\Omega$ is the orbital angular velocity), and far away from the merger and the resonant $f-$mode frequency, we can estimate the energy dissipated \textcolor{black}{in the star with mass $M_1$ as~\cite{Ghosh_2024}
\begin{equation}\label{eq:viscener}
    \Dot{E}_{\rm visc} = \frac{24\pi}{5}q^2(1+q)\frac{M_1^2}{R_1}\frac{Q_1^2}{\bar{\omega}_1^{4}}\left(\frac{R_1}{D}\right)^9\gamma ,
\end{equation}
where $M_1$ and $qM_1$ are the mass of the heavier star and its companion respectively with $q$ being the mass ratio ($q \leq 1$), $Q_1$ is the tidal coupling of the $l = m = 2$ $f$-mode, $R_1$ is the radius of the star 1, $\bar{\omega}_1$ the normalised (by $\sqrt{M_1/R_1^3}$) angular frequency of the $f$-mode and $D$ is the orbital separation. }

\textcolor{black}{\section{Microphysics of strange stars}}
\subsection{EoS for strange quark matter}
\label{subsec:EOS_QM}
We briefly describe the non-ideal bag model that we use to study unpaired strange quark matter ($u,d,s$) with electrons ($e$). The thermodynamic potential $\Omega_{\rm ni}$ (normalised by the volume) as a function of the particles' chemical  potential $\mu_i$ with $i=u,d,s,e$ and temperature $T$ can be written as follows
\begin{equation}
    \Omega_{\rm ni}=\Omega_{e}^{(0)}+a_4\sum_{f=u,d,s}\Omega_{f}^{(0)} +B_{\rm eff},
    \label{potentialnon-idealMIT}
\end{equation}
where $B_{\rm eff}$ is the bag constant, $a_4$ is a nonperturbative parameter, which induces a deviation of the ideal free Fermi gas expression, and $\Omega^{(0)}_{i}$ is the single contribution to the thermodynamic potential from the constituent particles described as ideal Fermi gases, which is expressed as follows~\cite{Wen_2005, PhysRevD.9.3320}
\begin{eqnarray}
    \Omega_{i}^{(0)}&=&-\frac{g_i T}{2\pi^2} \int_0^\infty k^2 dk \, \Bigg\{ \ln\left[1+\exp \left(-\frac{ E_{i,k}-\mu_i}{T} \right)\right]  \nonumber \\
    &+&\ln\left[1+\exp\left(-\frac{ E_{i,k}+\mu_i}{T} \right) \right] \Bigg\},
    \label{PotentialidealgasquarksfinT}
\end{eqnarray}    
where $E_{i,k}=\sqrt{k^2 +m_i^2}$, $g_i$ is the degeneracy factor and $m_i$ is the mass of the particles. For electrons, $g_e=2$, considers their spin and for quarks, $g_f=2N_c$, includes the spin and color charge degrees of freedom being $N_c=3$ the number of colors. The mass of the electron is around $0.511$ MeV, but here we consider it to be massless. In this approximation and following the procedure in Ref.~\cite{Laine_2016}, we get
\begin{equation}
    \Omega_e^{(0)}= -\frac{1}{12} \left(\frac{\mu_e^4}{\pi^2} +2\mu_e^2 T^2 +\frac{7}{15}\pi^2T^4 \right).
\label{PotentialidealgaselectronsfinT}
\end{equation}

In addition, for massive quarks at zero temperature, Eq.~\eqref{PotentialidealgasquarksfinT} simplifies as follows
\begin{eqnarray}
    \Omega^{(0)}_f&=&-\frac{N_c}{12\pi^2}  \left[ \mu_f k_f\left(\mu_f^2 -\frac{5}{2}m_f^2 \right) \right. \nonumber \\
    &+& \left. \frac{3}{2}m_f^4 \ln{\left( \frac{\mu_f+k_f}{m_f}\right)} 
    \right],
    \label{OmegaFermiquarks}
\end{eqnarray}
here $f=u,d,s$, $m_f$ denotes the quark mass, and $k_f \equiv \sqrt{\mu_f^2 -m_f^2}$.

For compact stars we impose the charge neutrality condition
\begin{equation}
    n_e+\frac{1}{3}n_s +\frac{1}{3}n_d =\frac{2}{3}n_u,
    \label{chargeneutral}
\end{equation}
and the beta equilibrium conditions
\begin{eqnarray}
    &&\mu_d =\mu_s, \\
    &&\mu_s=\mu_u +\mu_e.
\end{eqnarray}

With these constraints and defining the total baryon density
\begin{eqnarray}
    &&n_B \equiv \frac{1}{3} n_u +\frac{1}{3} n_d +\frac{1}{3} n_s,
\end{eqnarray}
we determine the chemical potentials and particle number densities at a fixed value of the baryon number density.

\begin{figure}[H]
\includegraphics[width=0.483\textwidth]{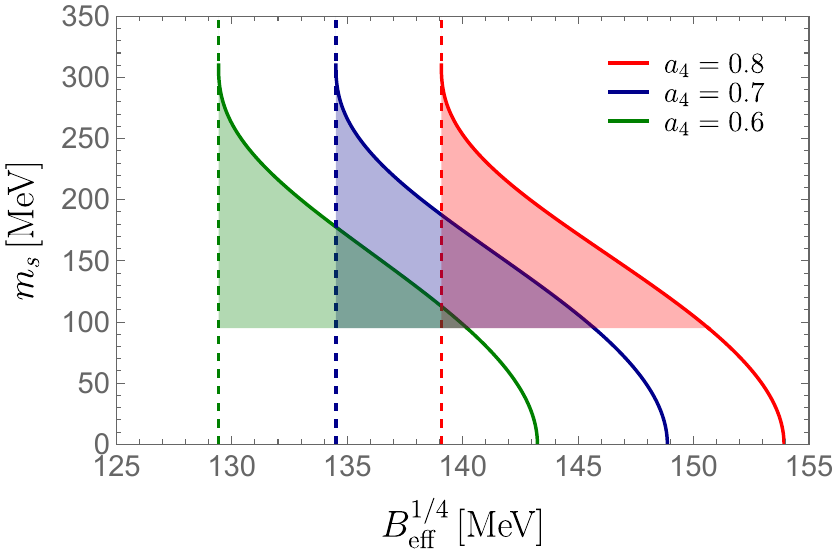}
\centering
\caption{Stability window for the non-ideal bag model. The strange quark mass $m_s$ is plotted as a function of $B_{\rm eff}^{1/4}$ for three different values of the nonperturbative parameter $a_4$. The shadow regions represent stable strange quark matter according to the Bodmer-Witten conjecture. \textcolor{black}{We have imposed $m_s=95$ MeV as the lowest value of the strange quark mass~\cite{ParticleDataGroup:2024cfk}.}}
\label{fig.stwindow}
\end{figure}

\subsection{Strange Star EoS and Stellar Properties}
\label{subsec.strangestarEoS}

Hereafter we study the parameter space allowed for the bag constant and the strange quark mass by the Bodmer-Witten conjecture, which states that strange quark matter at equal quark chemical potentials might be the true ground state of hadronic matter~\cite{PhysRevD.4.1601,PhysRevD.30.272}. In the zero temperature limit of Eq.\eqref{potentialnon-idealMIT}, we fix the bag constant so that the binding energy at zero pressure of strange quark matter is less than the energy per baryon of $^{56}\text{Fe}$ (i.e. $\varepsilon/n_B \leq 930\, \text{MeV}$) and also impose that two-flavour quark matter is not stable (i.e. $\varepsilon/n_B > 930\, \text{MeV}$), discarding the possibility of neutrons and protons decaying into up and down quarks~\cite{Lopes_2021,Torres_2013}. 

In Fig.~\ref{fig.stwindow}, we plot the regions which satisfy the constraints imposed by the Bodmer-Witten conjecture in the parameter space of $B_{\rm eff}^{1/4}$ and $m_s$ as reported in Ref.~\cite{Torres_2013}. The vertical lines represent the region where the energy density per baryon for two-flavour quark matter equals $930$ MeV and the other curved lines depict the values where the energy density per baryon for strange quark matter is $930$ MeV. The areas to the left of the vertical lines represent the regions where two-flavour quark matter is stable for a given value of $a_4$. The shaded regions show the stability windows setting $m_u=m_d=4$ MeV and $m_{s}=95$ MeV as the lowest value for the strange-quark mass as many results predicts approximately this value at a renormalization scale of $2$ GeV~\cite{ParticleDataGroup:2024cfk} and given that this mass is expected to be higher at densities of compact stars \textcolor{black}{as we will discuss below}. 

In Ref.~\cite{Fraga_2001} it was found that the perturbative QCD pressure up to $\mathcal{O}(\alpha_s^2)$ for three massless quarks can be approximated by a similar phenomenological bag model like the one in Eq.~\eqref{potentialnon-idealMIT} at chemical potentials from $300$ up to $650$ MeV relevant for quark stars by setting $a_4\approx0.63$ while the bag constant varies according to the renormalization scale choice. In this work, we set $a_4=0.8,\, 0.7,\, 0.6$, which induce small to medium deviations of an ideal free Fermi gas ($a_4=1$) and are close to the value reported in Ref.~\cite{Fraga_2001}. For the non-ideal bag model, we find that $m_{s,\text{max}}\approx 310$ MeV is the maximum value allowed for the strange quark mass consistent with the constraints mentioned above.
\begin{figure}[H]
\includegraphics[width=\linewidth]{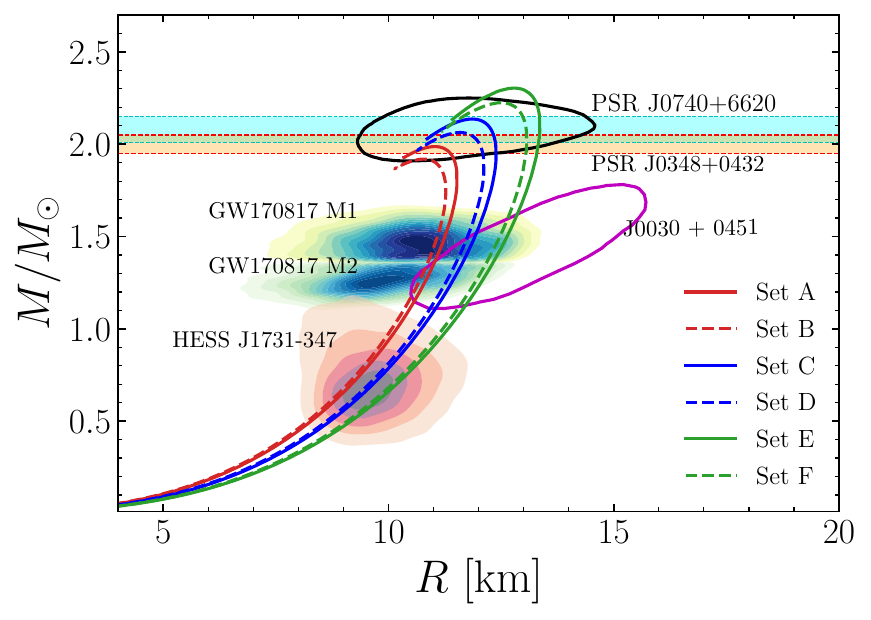}
    \centering
    \caption{M-R sequences for the EoS parametrization tabulated in~\cref{tab:EoSs}. Horizontal bands correspond to masses $M=2.08{\pm 0.07} M_{\odot}$ of PSR J0740$+$6620~\citep{Fonseca2021}  and $M=2.01^{+0.04}_{-0.04} M_{\odot}$ of PSR J0348$+$0432~\citep{Antoniadis2013}. The 90\% contour of M-R measurement for PSR J0740+6620 corresponding to Riley et al.~\citep{Riley2021,J0740_new} is shown in black, and for PSR  J0030+0451~\citep{Riley_2019,J0030_new} is shown in magenta. The M-R estimate of the lightest compact object HESS J1731-347 ~\citep{Hess} is shown by the shaded area labelled with `HESS J1731-347'. Although recent works~\citep{Hess_reanalysis} puts the
    original spectral analysis into question.  The M-R estimates of the two companion  stars of the merger event GW170817 are shown by the shaded area labelled with GW170817 M1 (M2).}
    \label{fig.mass-radius}
\end{figure}

According to the stability windows found in Fig.~\ref{fig.stwindow}, we define the sets of parameters ($m_s$, $a_4$ and $B_{\rm eff}^{1/4}$) in~\cref{tab:EoSs} so as to determine the EoSs, while assuming $m_u=m_d=4$ MeV and massless electrons. A variation of the bag constant in the allowed domain does not have a strong impact on the quantities we are interested\textcolor{black}{. I}n particular the bulk viscosity does not depend on this value. Regarding the strange quark mass, we choose $m_s=150,\,200$ MeV given that it is expected to be higher than the values reported in Ref.~\cite{ParticleDataGroup:2024cfk} at densities a few times the nuclear saturation density, due to the presence of quark antiquark condensates, as it occurs in Nambu-Jona-Lasinio models~\cite{BUBALLA_2005}.   Then, we solve the Tolman-Oppenheimer-Volkoff equations (TOV equations)
to obtain the mass (M)- radius (R) relations and the $f$-mode oscillation frequencies within the Cowling approximation as described in the~\cref{app:f_mode}. In Fig.~\ref{fig.mass-radius}, we display the M-R relations obtained with the set of parameters described above. Comparing the curves obtained using sets A-B, C-D, and E-F, where $a_4$ and the bag constant are fixed, we infer that decreasing the value of the strange quark mass produces higher maximum masses and radii. Using the EoS of the non-ideal bag model it is possible to produce maximum masses above $2M_\odot$ unlike the original MIT bag model, which generates at most a maximum mass up to $1.85M_{\odot}$~\cite{Lopes_2021}. A similar effect that is more remarkable occurs when we decrease $a_4$~\cite{Lopes_2021} (see, for example, set A and C, the small variation of the bag constant does not have a significant effect).

\begin{table}[H]
\begin{center}
\begin{tabular}{| c | c | c | c |}
\hline
 Set  & $m_s[{\rm MeV}]$ & $a_4$ & $B_{\rm eff}^{1/4}[{\rm MeV}]$ \\  \hline
A &$150$ & $0.8$ & $140$ \\
B &$200$ & $0.8$ & $140$ \\
C &$150$ & $0.7$ & $135$ \\
D &$200$ & $0.7$ & $135$ \\
E &$150$ & $0.6$ & $130$ \\
F &$200$ & $0.6$ & $130$ \\ \hline
\end{tabular}
\caption{Sets of allowed EoS parameters \textcolor{black}{($m_s$, $a_4$ and $B_{\rm eff}^{1/4}$)} subject to the Bodmer-Witten conjecture.}
\label{tab:EoSs}
\end{center}
\end{table} 
It is evident that the considered EoS parametrization successfully satisfies key astrophysical observational constraints. The EoS models not only explain the maximum observed pulsar mass but also align well with the M-R posterior distributions of \textcolor{black}{PSR J0740+6620~\citep{Riley2021,J0740_new} and PSR J0030+0451~\citep{Riley_2019,J0030_new}}. For the GW170817 merger event, the equivalent M-R posteriors of the component stars, as provided in \citep{Abbott2018}, are derived based on either neutron star EoS models, particularly the piecewise polytropic spectral decomposition approach, or EoS-independent universal relations that were specifically developed for neutron stars. However, in Fig.~\ref{fig.mass-radius}, we present instead the equivalent strange star M-R posteriors, obtained using the mass and tidal deformability measurements for GW170817 from~\citep{Abbott2019} in conjunction with the use of universal relations specifically developed for strange quark stars~\citep{Yagi2017}. Additionally, we have included the M-R posterior of the low-mass compact object HESS J1731-347~\citep{Horvath2023}, which was initially proposed as a strong candidate for a strange star.

In~\cref{tab:f-modeMModot}, we show the values of $f$-mode frequencies $\omega/2\pi$ in kilohertz for the EoSs defined in~\cref{tab:EoSs} varying the masses of the stars $M/M_{\odot}$.

\begin{table}[H]
\begin{center}
\begin{tabular}{| c | c | c | c | }
\hline
 Set & $\omega/2\pi \,[{\rm kHz}]$ & $\omega/2\pi \,[{\rm kHz}]$ &$\omega/2\pi\, [{\rm kHz}]$ \\
 & $(M/M_\odot =1.4)$ & $(M/M_\odot =1.6)$ & $(M/M_\odot =1.8)$ \\ \hline 
A & $2.291$ & $2.306$ & $2.343$ \\ 
B & $2.333$ & $2.356$ & $2.413$ \\ 
C & $2.127$ & $2.135$ & $2.154$ \\ 
D & $2.164$ & $2.177$ & $2.205$ \\ 
E & $1.970$ & $1.974$ & $1.985$ \\ 
F & $2.003$ & $2.010$ & $2.026$ \\ \hline
\end{tabular}
\caption{$f$-mode frequencies $\omega/2\pi$ in kHz at $M/M_\odot =1.4,\,1.6,\,1.8$ for the sets defined in~\cref{tab:EoSs}.}
\label{tab:f-modeMModot}
\end{center}
\end{table}  


\subsection{Bulk viscosity in strange quark stars}
\label{subsec:bulkviscosity}

To complete this section, we address the bulk viscosity induced by the electroweak force in unpaired three-flavor quark matter following the formalism in Ref.~\cite{PhysRevD.109.123022}. 

Tidal perturbations induce volume expansion and compression of the medium, which subject the system to instantaneous deviations from chemical equilibrium. For three-flavour quark matter in the normal phase, where the medium is transparent to neutrinos, the non-leptonic weak process
\begin{equation}
    u+d \leftrightarrow u+s
\label{EW-non-processes}
\end{equation}
provides the main contribution to dissipate energy and re-establish chemical equilibrium at low temperatures ($T\leq 10^{10}$ K)~\cite{Sad:2007afd,Alford_2010,Cruz_Rojas_2024,PhysRevD.109.123022}. This temperature regime is relevant for the tidal heating in strange quark stars.

At linear order in the chemical imbalance $\mu_1=\mu_s-\mu_d$, the rate difference associated to the processes in Eq.~\eqref{EW-non-processes} can be expressed as
\begin{equation}
\Gamma_{s+u \rightarrow d+u} - \Gamma_{d+u \rightarrow s+u}=\mu_1 \lambda_1. \label{rate1}  
\end{equation}

At tree level in the low-temperature limit this rate difference has been computed assuming a small mismatch between the down and strange quark chemical potentials~\cite{SAWYER1989412}, so that
\begin{eqnarray}
    \lambda_1 &=& \frac{1}{8\pi^3}G_F^2 \sin^2 \Theta_C \cos^2 \Theta_C T^2 \bigg\{ \frac{3}{5}[32 k_u^5 - (k_d-k_s)^5] \nonumber \\
    &-& [8k_u^3 -(k_d-k_s)^3] (4k_u^2 +k_s^2 +k_d^2 +2k_d \mu_s) \nonumber \\
    &+&12(2k_u-k_d+k_s)k_u^2 (k_s^2+k_d^2+2k_d \mu_s)\bigg\}, \label{lambda1}
\end{eqnarray}
where $G_F=1.166\times 10^{-5}\, \text{GeV}^{-2}$ is the Fermi coupling constant, $\Theta_C$ is the Cabibbo angle, and $k_f \equiv \sqrt{\mu_f^2 -m_f^2}$. 
The expression in Eq~\eqref{lambda1} considers part of the quark mass dependence of the rates. However, for high values of the strange quark mass the rate difference is a complex expression that can only be computed numerically~\cite{Madsen:1993xx,Anand_1997}. 
Further, Eq.~\eqref{lambda1} expanded for $m_s=0$ and for massless light quarks at
$\mu_d=\mu_u$ allow us to get the well-known result~\cite{Madsen:1993xx} 
\begin{equation}
\lambda_1\approx\frac{64}{5\pi^3}G_F^2 \sin^2{\Theta_C}\cos^2{\Theta_C}\mu_u^5 T^2.
\end{equation}
Considering the conditions described above, bulk viscosity at first-order hydrodynamics can be written as
\begin{equation}
    \zeta=\frac{\lambda_1 C_1^2}{(A_{dd}+A_{ss})^2\lambda_1^2 +\omega^2},
    \label{eq.bulkviscosity}
\end{equation} 
here $\omega$ is the angular frequency of the perturbation (the $f$-mode angular frequency in this work) and we define  
\begin{eqnarray}
&&C_1 \equiv n_{s,0}A_{ss}-n_{d,0}A_{dd},
\end{eqnarray}
where $n_{j,0}$ are the instantaneous values of the particle number densities in chemical equilibrium and $A_{ij}$ are the susceptibilities of the constituent particles written in terms of the partial derivatives of the particle's chemical potentials $\mu_i$ with respect to the particle number density $n_i$:
\begin{equation}
A_{ij}=\left(\partial \mu_i/\partial n_j \right)_{n_{k\neq j},T}.  
\end{equation}
It is worth to note that the result in Eq.~\eqref{eq.bulkviscosity} only reflects the diagonal contribution of the susceptibilities (see Ref.~\cite{Cruz_Rojas_2024} where non-diagonal susceptibilities are considered) and that the temperature dependence enters via the rates, as we neglect the small thermal corrections in the EoS. This approximation might not be valid for temperatures larger than $10^{10}$ K, as temperature corrections to the EoS might become relevant. 
\begin{figure}[H]
    \centering
    \includegraphics[width=0.5\textwidth]{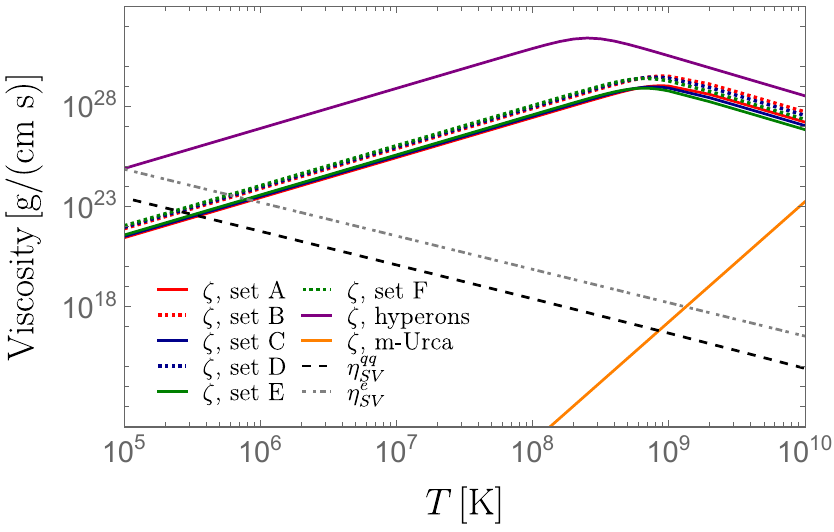}
    \caption{Relative strengths of various sources of viscosities of dense matter inside neutron stars as a function of temperature at  $n_B=2.5n_0$ with $n_0\approx 0.15$ fm$^{-3}$. The bulk viscosity at $\omega= 2\pi$ kHz before and after the resonant peak is shown for the strange-quark-matter EoSs (Set A-F) detailed in~\cref{tab:EoSs}, non-leptonic reactions involving $\Lambda$ hyperons, and m-Urca processes in nuclear matter. We also plot, the standard neutron matter shear viscosity coming from $ee$ scattering $\eta_{SV}^{e}$, and the shear viscosity in degenerate quark matter coming from quark-quark scattering $\eta_{SV}^{qq}$.}
    \label{fig:BV_temp}
\end{figure}

In Fig.~\ref{fig:BV_temp}, we display the bulk viscosity as a function of temperature at the typical characteristic value of oscillation frequency of $1$ kHz~\cite{Pradhan_2021} and $n_B=2.5n_0$, being $n_0$ the nuclear saturation density. First, we plot the bulk viscosity employing the EoSs described in~\cref{tab:EoSs}. Further, we include the bulk viscosity from non-leptonic reactions involving $\Lambda$-hyperons~\citep{Ghosh_2024} as well as modified-Urca (m-Urca) reactions to compare their relative strengths. Lastly, we plot the neutron matter shear viscosity from electron scattering $\eta_{SV}^{e}$ according to~\cite{Alford_2012sv} at a proton fraction $x_p=0.1$ and $3n_0$ and the shear viscosity in degenerate quark matter from quark-quark scattering $\eta_{SV}^{qq}$ at approximated equal quark chemical potentials $\mu_q=350$ MeV obtained at $2.5n_0$ and $\alpha_s\approx 0.7$~\cite{PhysRevD.48.2916}. At $T\sim 10^9$ K, bulk viscosity in unpaired strange quark matter reaches its resonant maximum value ($\sim 10^{29}$ g cm$^{-1}$s$^{-1}$), which is several orders of magnitude higher than the $ee$ shear, m-Urca bulk viscosity~\cite{SAWYER1989412} and the $qq$ shear viscosity in this temperature regime. Regarding the bulk viscosity in strange quark matter, the values associated with the sets B, D and F (for which $m_s=200$ MeV) are a few times higher than the ones of sets A, C and E (with $m_s=150$ MeV). This trend is expected given the strong $m_s$-dependence of $\zeta$ in Eq.~\eqref{eq.bulkviscosity}. On the other hand, the bulk viscosity does not depend on the bag constant since in this model there is neither temperature nor chemical potential dependence on $B_{\rm eff}$. Finally, its dependence on $a_4$ is almost negligible below $T=10^9$ K and small for higher temperatures up to $10^{11}$ K.

\section{Temperature estimate during binary inspiral}
\label{sec:Temp}

Since the dissipated tidal energy is converted to thermal energy and heats up the system during the inspiral, it is crucial to obtain an estimate of the average temperature we expect from this tidal dissipation. \textcolor{black}{In this study, we consider equal mass binaries, and hence we remove the subscript associated with each star from here on. Considering the high thermal conductivity, the stars are expected to be isothermal during the inspiral. Considering the effects of General Relativity, the constant redshifted temperature, $\tilde{T}$ relates to the local temperature, $T$ by 
\begin{equation}
    \tilde{T} = Te^{\nu},
\end{equation}
where $\nu$ is the metric function (see Eq.~\eqref{eq.gravphase}), which determines the gravitational redshift.} During the binary inspiral, \textcolor{black}{the thermal evolution for the redshifted temperature} of the strange quark star can be written as~\cite{Lai_1994}
\begin{equation}\label{eq:evol_T}
    \frac{dU}{dt} = \Dot{E}_{\rm visc} + \Dot{E}_{\rm cool},
\end{equation}
where \textcolor{black}{$U$ is the internal energy of the system,} $\Dot{E}_{\rm cool}$ denotes the rate of cooling due to neutrino emission and surface photon luminosity. As shown in Appendix~\ref{app:neutrino_emissivity}, \textcolor{black}{the cooling timescale due to neutrino emission from strange quark stars is much longer} than the inspiral timescale (of the order of $\sim 100$ s) for the relevant range of temperatures $\leq 10^9$ K. Therefore, the cooling due to neutrino emission can be neglected in this calculation. In addition, photon emission in quark stars has been studied in Ref.~\cite{Cheng_2003}. \textcolor{black}{This results to be smaller than the blackbody radiation and not relevant for the temperatures we consider in this work. Thus}, the rate of change of the internal energy is given by
\begin{equation}\label{eq:evol_T_CV}
    \frac{dU}{dt} = C_V \frac{d\tilde{T}}{dt},
\end{equation}
where $C_V$ is the \textcolor{black}{average heat capacity of strange quark matter. This quantity can be computed over the entire stellar volume following a similar reasoning as in Ref.~\cite{Ofengeim:2017xxr}
\begin{equation}
    C_V=\int_0^{R} c_V(\rho,\tilde{T})\frac{4\pi r^2 dr}{\sqrt{1-2m/r}}.
\end{equation}
Here $R$ is the radius of the star as stated above, and $m=m(r)$ is the gravitational mass inside a sphere with circumferential radius $r$.}

\textcolor{black}{The specific heat capacity, $c_V$ can be computed in strange quark matter using the following thermodynamic relation~\cite{Ferrer_2021} 
\begin{equation}
    c_{V}=-T\frac{\partial^2\Omega_{\rm ni}}{\partial T^2}\Bigg|_{\mu_i},
\end{equation}
with $i=u,d,s,e$. Hereafter, we use the low-temperature approximation $T<\mu_i$ and $m_i<\mu_f$. These approximations are valid for the chemical potentials and temperatures considered for electric charge neutral strange quark matter in chemical equilibrium. Following a similar procedure as the one to obtain Eq.~(11) in Ref.~\cite{Ferrer_2021}, we get the specific heat capacity of strange quark matter
\begin{equation}
    c_{V} 
    = \Tilde{T}e^{-\nu}\left(\frac{\mu_e^2}{3}+a_4\sum_{f=u,d,s}\mu_f\sqrt{\mu_f^2-m_f^2}\right).
\end{equation}
Note that with the definitions above and considering a constant redshifted temperature over the stellar volume, we can express the average heat capacity as follows
\begin{equation}\label{eq.spheatcap}
    C_{V}=\sigma\tilde{T} ,
\end{equation}
where $\sigma$ depends on the variation of the chemical potentials, the gravitational redshift $\nu$ profile, and also the mass configurations.}

\textcolor{black}{In~\cref{tab:sigma} we provide the values of $\sigma$ using the non-ideal bag model for the different equal-mass configurations considered in this work. }
\begin{table}[H]
\begin{center}
\begin{tabular}{| c | c | c | c | }
\hline
 Set & $\sigma$ $\left[\text{erg}/\text{K}^2 \right]$ & $\sigma$ $\left[\text{erg}/\text{K}^2 \right]$ &$\sigma$ $\left[\text{erg}/\text{K}^2 \right]$ \\
 & $(M/M_\odot =1.4)$ & $(M/M_\odot =1.6)$ & $(M/M_\odot =1.8)$ \\ \hline 
A & $9.641 \times 10^{29}$ & $1.125 \times 10^{30}$ & $1.295 \times 10^{30}$ \\ 
B & $9.578 \times 10^{29}$ & $1.118 \times 10^{30}$ & $1.288 \times 10^{30}$ \\ 
C & $9.615 \times 10^{29}$ & $1.119 \times 10^{30}$ & $1.283 \times 10^{30}$ \\ 
D & $9.544 \times 10^{29}$ & $1.112 \times 10^{30}$ & $1.275 \times 10^{30}$ \\ 
E & $9.517 \times 10^{29}$ & $1.108 \times 10^{30}$ & $1.267 \times 10^{30}$ \\ 
F & $9.466 \times 10^{29}$ & $1.099 \times 10^{30}$ & $1.257 \times 10^{30}$ \\ \hline
\end{tabular}
\caption{\textcolor{black}{$\sigma$ values in the average heat capacity ($C_V = \sigma\tilde{T}$) for different masses and EoSs considered in this work.}}
\label{tab:sigma}
\end{center}
\end{table}  

\textcolor{black}{Given the local-temperature} dependence of the bulk viscosity in unpaired strange quark matter described in Sec.~\ref{subsec:bulkviscosity} (see also Eq.~\eqref{lambda1}), the bulk viscous dissipation rate can be parametrized as (see Eq.~\eqref{eq:gammabulk})
\begin{equation}
    \gamma = \frac{A\tilde{T}^2}{B+\tilde{T}^4}.
\end{equation}
\textcolor{black}{The parameters $A$ and $B$ are numerically fitted to the $\gamma$ profiles with the isothermal temperature corrected for the gravitational redshift.}
\begin{figure}[H] 
\includegraphics[width=0.483\textwidth]{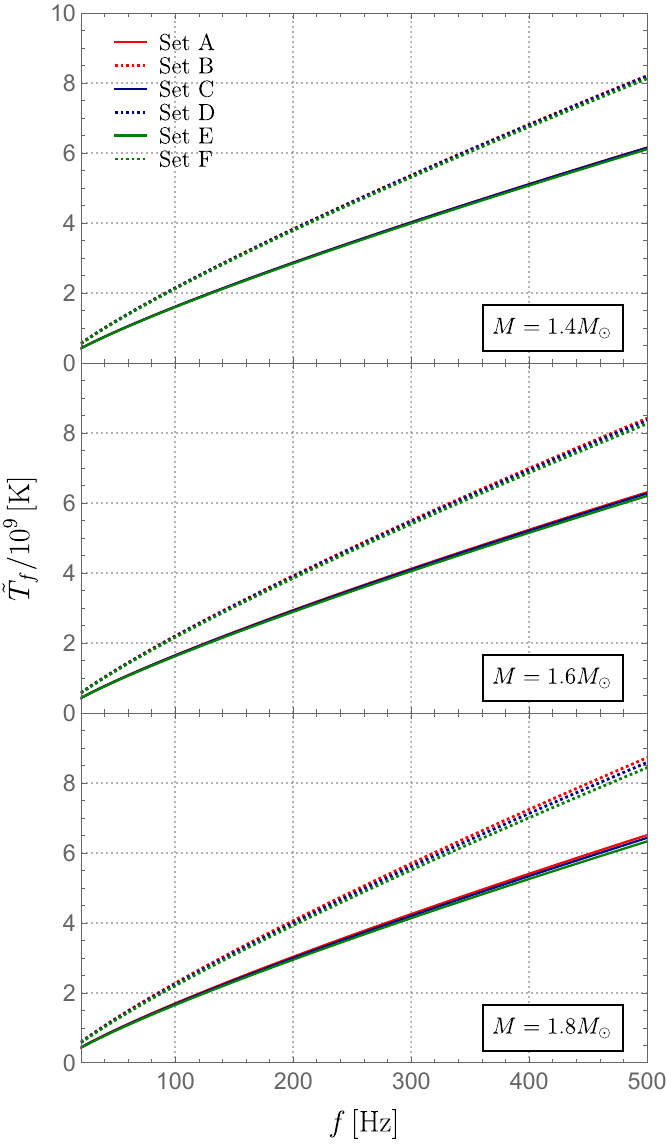}
\caption{Estimates of the average \textcolor{black}{final} temperature due to the tidal dissipation as a function of GW frequency for the six different EoSs (Sets A-F) considered for three different equal mass binary systems: a) $1.4M_{\odot}$ (top panel) b) $1.6M_{\odot}$ (mid panel) and c) $1.8M_{\odot}$ (bottom panel).}
\label{fig:temperature}
\end{figure}
\textcolor{black}{Then, we obtain an estimate of the temperature reached, $\tilde{T}_f$, as a function of the separation of the stars $D$. Using Eq.~\eqref{eq:viscener} with the orbital evolution equation  
\begin{equation}
    \frac{dD}{dt} = -\frac{64}{5}\frac{M_1^3q(1+q)}{D^3},
\end{equation}
we integrate the thermal evolution equation in Eq.~\eqref{eq:evol_T} 
from $D \to \infty$, when the stars are far apart and at a very low temperature $\tilde{T}_0 \sim 10^5 - 10^6$ K, to finally obtain}
\textcolor{black}{
\begin{equation}\label{eq:final_T}
        \frac{\tilde{T}_f^4}{4} + B\ln{\left(\frac{\tilde{T}_f}{\tilde{T}_0}\right)}  = \frac{\tilde{T}_0^4}{4} + \frac{\pi}{648}\frac{Q^2q}{\bar{\omega}^{4}} \frac{A}{\sigma}\frac{R^3}{M_1}   \left(\frac{3R}{D}\right)^5.
\end{equation}
}
Fig.~\ref{fig:temperature} shows the estimate of the average temperature, $\tilde{T}_f$, of the star for the EoSs considered in this study and for three different equal mass binary neutron star systems. For all scenarios considered, we found temperatures of a few times $10^{9}$ K reached at the end of the inspiral with a slight increase towards high mass binaries. In addition, as the strange quark mass increases from $150$ MeV (sets A, C and E) to $200$ MeV (sets B, D and F) the final temperature also increases around \textcolor{black}{$33 \%$} for the three stellar configurations. This is because the bulk viscosity increases with the strange quark mass. Moreover, we see a slight decrease in the temperature as we decrease $a_4$ and $B_{\rm eff}$ at fixed strange quark mass.

\section{Phase Contribution to gravitational waveform}
\label{sec:phase}
During the early stages of a binary neutron star evolution away from the merger, the change in the orbital frequency due to the emission of GWs is much smaller than the orbital frequency itself. In this regime, the evolution of the orbital phase $\Phi(t)$ of the binary system is computed as a perturbative expansion in a small parameter, typically taken to be the characteristic velocity $v = (\pi M f)^{1/3}$,$M$ being the total mass of the binary and $f = \Omega/\pi$ being the GW frequency. This analytical procedure requires $v\ll 1$, which makes it useful in the early inspiral phase of the binary. In this adiabatic regime, the loss of binding energy $E(v)$ of the two-body system with time equals the GW flux emitted to future null infinity $\mathcal{F}^{\infty}(v)$ plus the energy dissipated to thermal energy due to the viscous dissipation $\dot{E}_{\rm visc}(v)$. Then, the energy balance condition becomes
\begin{equation}\label{eq:energybalance}
    -\dv{E(v)}{t}=\mathcal{F}^{\infty}(v)+\dot{E}_{\rm visc}(v).
\end{equation}

The evolution of the orbital phase $\Phi$ and the characteristic velocity $v$ are obtained from the following equations~\citep{TaylorF2}
\begin{equation}
    \dv{\Phi}{t} = \frac{v^3}{M},
\end{equation}
and
\begin{equation}
\dv{v}{t}=-\frac{\mathcal{F}(v)}{E'(v)},
\end{equation}
where $E'(v)=dE(v)/dv$ and $\mathcal{F}(v)=\mathcal{F}^{\infty}(v)+\dot{E}_{\rm visc}(v)$. In the frequency domain using the stationary
phase approximation (SPA), the gravitational waveform can be written as
$\Tilde{h}(v) =  \Tilde{A}(v) e^{-i\Phi(v)}$~\cite{Tichy:1999pv,TaylorF2} with
the phase given as
\begin{equation}
\label{psi}
    \Phi(v)=\frac{2t_c v^3}{M} - 2\phi_c - \frac{\pi}{4} - \frac{2}{M}\int (v^3-\Bar{v}^3)\frac{E'(\Bar{v})}{\mathcal{F}(\Bar{v})}\dd\Bar{v},
\end{equation}
where $\phi_c,\,t_c$ are constants. The separation $D$ between the stars in Eq.~\eqref{eq:viscener} and the orbital frequency $\Omega$ are related by
\begin{equation}
    \Omega^2=\frac{M}{D^3}\,.
\end{equation}
Employing the previous relations given for the characteristic velocity and the GW frequency, we get the separation distance as a function of $v$ as follows
\begin{equation}
    D(v)=\frac{M}{v^2}\,.
\end{equation}
Substituting this result  in Eq.~\eqref{eq:viscener}, we have
\begin{equation}\label{eq:ener_v}
    \dot{E}_{\rm visc} = \frac{24\pi}{5} q^2(1+q)\frac{(M/2)^2}{R}\frac{Q^2}{\bar{\omega}^4}\left(\frac{R}{M}\right)^9 \gamma (v)\, v^{18}\,.
\end{equation}
Note that we express $\gamma$ as a function of the velocity simply because the magnitude of the viscosity strength depends strongly on the temperature, and due to viscous dissipation we expect the star also to heat up during inspiral. Also, for equal mass binary systems considered in this work, we set $q =1$ and also multiply the expression in Eq.~\eqref{eq:ener_v} by a factor of $2$ to consider the contribution of the both the binary components. 

Up to the leading order (LO) and next-to leading order (NLO), the post-Newtonian expansion for the functions $E(v)$ and $\mathcal{F}^{\infty}(v)$   have the general form~\cite{Isoyama:2017tbp}
\begin{equation}\label{EV}
    E(v) = -\frac{1}{2}\eta M v^2\left[1 - \frac{9+\eta}{12}v^2\right]\,,
\end{equation}
and 
\begin{equation}\label{FV}
    \mathcal{F}^{\infty}(v) = \frac{32}{5}v^{10}\eta^2\left[1 - v^2\left(\frac{1247}{336}+\frac{35\eta}{12}\right) + 4\pi v^3\right],
\end{equation}
where \textcolor{black}{$\eta =M_1M_2/M^2$} is the symmetric mass ratio. For equal mass systems $\eta = 1/4$. Plugging the expressions for $\Dot{E}_{\rm visc}$, $E(v)$ and $\mathcal{F}^{\infty}(v)$  from Eqs.~\eqref{eq:ener_v},~\eqref{EV} and~\eqref{FV} respectively in Eq.~\eqref{psi}, we numerically \textcolor{black}{integrate it} to obtain the phase of gravitational waves. To compute the additional phase due to viscous dissipation $\Delta\Phi$, we simply subtract this phase from the same obtained without considering the contribution from viscous dissipation $\dot{E}_{\rm visc}$. \textcolor{black}{If we ignore the additional frequency dependence of the dissipation rate $\gamma$, analytic computation of Eq.~\eqref{psi} tells us that tidal dissipation effects enter the GW phase at 4th Post Newtonian (PN) order compared to the point-particle phase with an additional $\log(v)$ term making it non-degenerate with any other physical binary parameters, particularly time of coalescence~\cite{Ripley_2023}}. \\
\begin{figure}[H] 
\includegraphics[width=0.483\textwidth]{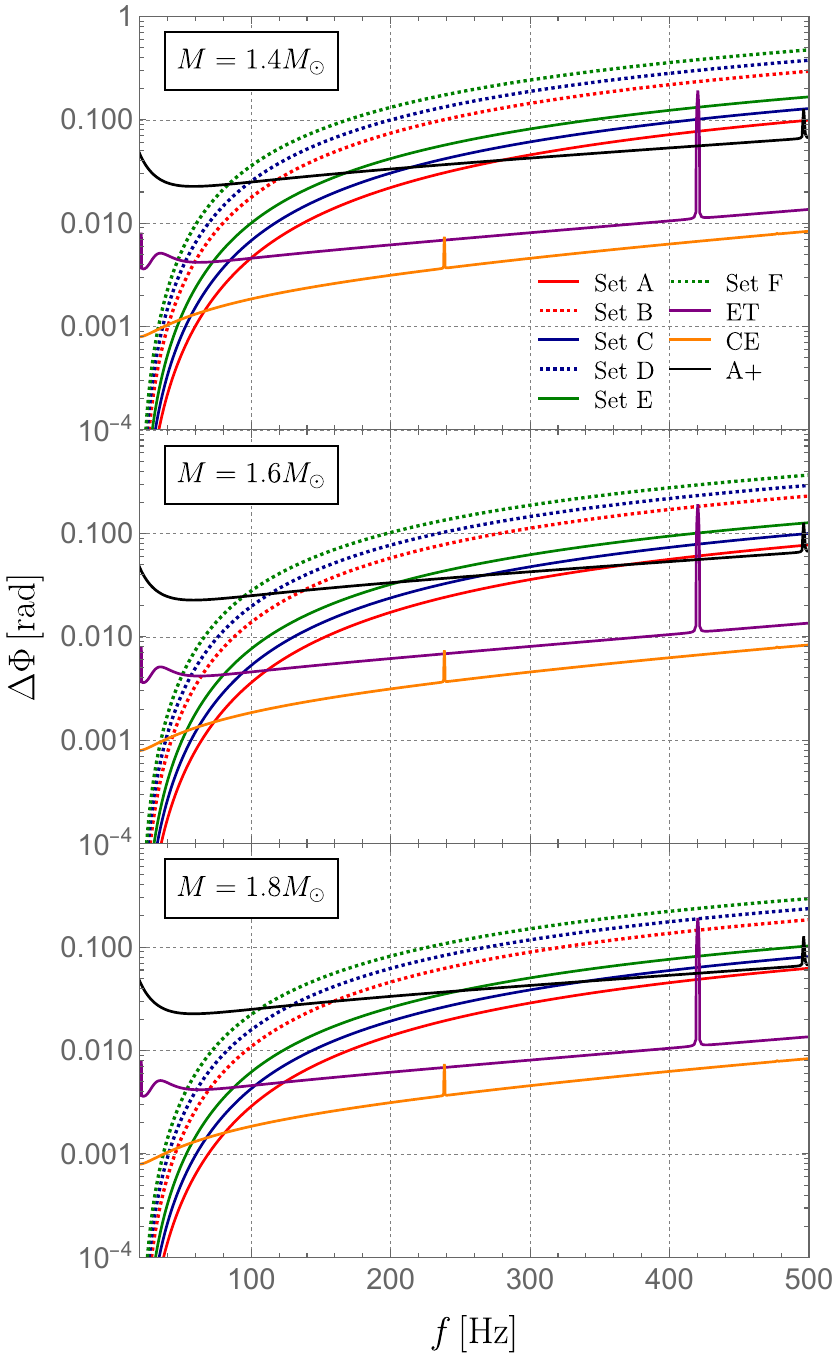}
\caption{The cumulative phase due to the tidal dissipation as a function of GW frequency for the sets of EoSs and three different equal mass binary systems: a) $1.4M_{\odot}$ (top panel) b) $1.6M_{\odot}$ (mid panel) and c) $1.8M_{\odot}$ (bottom panel). The phase uncertainty limits for next generation detectors : A+~\citep{Aplus_sensitivity}, 40-km Cosmic Explorer(CE)~\citep{CE,CE2} and Einstein Telescope(ET)~\citep{Hild_2011} are also plotted in solid lines for a source at $100$ Mpc.}
\label{fig:phase}
\end{figure}
In Fig.~\ref{fig:phase} we show this additional phase for the EoSs considered in this study and three different BNS systems of equal mass. \textcolor{black}{For all these systems considered, the additional phase obtained is of the order of $\sim 0.1 - 0.5$ rad for the sets with $m_s = 200$ MeV (sets B, D and F) and $0.06-0.2$ rad for the sets with $m_s = 150$ MeV (sets A, C and E).} As seen in the case of the temperature estimates as well as in Sec.~\ref{sec:TIDALDISSIPATEDENERGY}, this is because the rate of dissipated energy is proportional to the bulk viscous dissipation rate and therefore to the bulk viscosity. Thus, we expect that $\dot{E}_{\rm visc}$ increases as the strange quark mass increases. \textcolor{black}{We also see a slight increase in the phase difference towards low mass binaries for all the EoSs} and also for decreasing $a_4$ and bag constant $B_{\rm eff}^{1/4}$ for all the equal mass binaries.

As discussed in ~\citep{read2023waveform}, for this phase shift in the gravitational waveform to be detectable by current or next generation ground-based GW detectors, it should be \textcolor{black}{larger} than the tolerance limit
\sg{
\begin{equation}\label{eq:phase_uncertainty}
    |\Delta \Phi (f)| \geq \frac{\sqrt{S_n(f) }}{2\tilde{A}(f)\sqrt{f}}~,
\end{equation}
}

where $S_n(f)$ is the power spectral density of the noise for each detector and \sg{$\tilde{A}(f)$} is the gravitational wave amplitude of the waveform.  Current waveform uncertainties at frequencies much lower than the merger frequency are much lower than the estimated phase uncertainties we are interested in here. So, we only consider the amplitude from a single state-of-the-art waveform model `IMRPhenomPv2\_NRTidalv2' for this analysis ~\citep{Ajith_2007,Husa2016,Khan2016,Hannam2014,Tim2017,Tim2019}. In Fig.~\ref{fig:phase}, we plot this estimated phase sensitivity according to Eq.~\eqref{eq:phase_uncertainty} for the detector sensitivity of LIGO A+~\citep{Aplus_sensitivity}, 40-km Cosmic Explorer (CE)~\citep{CE,CE2} and the Einstein Telescope (ET)-D~\citep{Hild_2011}, at a distance
of 100 Mpc. From these estimates, we see that this extra phase shift should in principle be detectable with next-generation detectors for nearby sources with high enough signal-to-noise (SNR) ratio.

\section{Discussion}
\label{sec:discussion}
In this work we have investigated the effect of tidal dissipation on GW emission during the binary inspiral of strange quark stars. Although viscous dissipation was previously thought to be negligible for the GW inspiral phase, recent studies involving hyperons~\citep{Ghosh_2024} have shown that if the bulk viscosity is strong enough at low temperatures ($T \ll 10^{10}$ K), it can produce an effect during the inspiral detectable using future generation GW detectors with increased sensitivity. Deconfined strange quark matter that can be a stable component in the high density core of neutron stars also produces high bulk viscosity that reaches its maximum at low temperatures ($\sim 10^9$ K). 

We have considered the Newtonian tidal interaction of the binary components as linear perturbations of a background solution for a star. We have analysed the tidal energy dissipated in this bulk-viscous medium from the fundamental $f-$mode of the star taking into account its dominant coupling with the external tidal field. Since viscosity coefficients are dependent on the stellar temperature, the estimate of the energy dissipation is correlated to the accurate estimation of the temperature during the inspiral. Thermal evolution of the system during the inspiral has been determined considering the dissipated energy and the cooling of the system due to neutrino emission. Regarding the latter, we have computed the cooling timescale employing the neutrino emissivity in unpaired strange quark matter~\cite{PhysRevD.70.114037}, this being much longer than the inspiral timescale, so it will not have any impact on the temperature estimates.

In order to model unpaired strange quark matter, we have resorted to the non-ideal bag model. We have considered six sets of EoSs with varying strange quark mass $m_s$ and the bag constant $B_{\rm eff}$ that span the current EoS uncertainty range and satisfy the astrophysical constraints of maximum observed mass and tidal deformability from GW170817. We have also calculated the \textcolor{black}{average} heat capacity of quark matter for the EoSs based on the non-ideal bag model to accurately determine the temperature reached at the end of the inspiral. For all the EoSs and equal mass binary systems (of $1.4M_{\odot},1.6M_{\odot}$ and $1.8M_{\odot}$) considered, we have found temperatures of few times $10^{9}$ K reached at the end of the inspiral. These estimates are of the same order as those obtained from hyperonic bulk viscous dissipation~\citep{Ghosh_2024} and a few orders of magnitude higher than shear viscosity from $nn$ or $ee$ scattering~\citep{Lai_1994} and bulk viscosity from Urca reactions~\citep{Arras2019}. 
However, the temperatures are not high enough to require the inclusion of thermal effects in the EoS during the inspiral. 

Further, we have used this viscous dissipated energy to estimate the additional phase contribution to the frequency domain gravitational waveform using the single phase approximation. \textcolor{black}{For all the equal mass binary systems considered, the additional phase obtained is $\sim 0.06-0.2$ rad and $\sim 0.1 - 0.5$ rad for the sets where $m_s=150$ and $200$ MeV, respectively}. Since it is expected that nonperturbative effects at high densities in QCD can contribute to make the strange quark mass higher, further constraints on this parameter and their impact on the bulk viscosity calculation are a line of improvement in the estimation of the phase shift in the gravitational waveform. 

We have found that this extra phase shift in the gravitational waveform is higher than the expected upper limits of the phase uncertainty for nearby sources (at distances of $100$ Mpc) considering current state-of-the-art waveform models and the proposed future generation GW detector sensitivities. Therefore, this additional phase should in principle be detectable with next generation detectors for high SNR events. It should be noted that the inferred phase uncertainties only serve as approximate upper limits of this effect and should be incorporated into waveform models to obtain accurate estimates. When compared with the estimates for hyperonic bulk viscous dissipation, we get a detectable phase difference for $1.4M_{\odot}$ and $1.6M_{\odot}$ binaries of strange quark stars.  For hyperons, this happens only for very massive systems ($\geq 1.8M_{\odot}$), \textcolor{black}{as the size of hyperon core increases with the mass of neutron stars}. Therefore, the signature of tidal dissipation during binary inspiral in low mass binary systems ($\sim 1.4M_{\odot}$) can be a `smoking gun' signature for the strange quark stars. 

There are several future directions that can be explored based on this work. Firstly, more consolidated efforts should be given to develop state-of-the-art gravitational waveform models that incorporate the dissipative aspects of tides. Recent works have incorporated this dissipative effect in frequency domain models~\citep{ghosh2025tidaldissipationbinaryneutron} or in terms of `tidal lag' in effective theory~\citep{Ripley_2023,Ripley_2024,Saketh_2024}, but these studies should be improved to accurately capture the tidal dissipation over the entire parameter space of mass and frequency during inspiral. Secondly, it would be interesting to determine the amount of dissipation in hybrid stars as well, although we expect this to be much smaller than strange stars given the smaller quark core. Finally, it is also instructive to consider other phases of strange quark matter, whether in the core of a hybrid star or as strange quark star. For example, the bulk viscosity has already been studied for color-superconducting strange quark matter in the 2SC phase~\cite{Alford_2006,Alford_2024}, spin-one color-superconducting strange quark matter~\cite{Wang_2010}, and color-flavor-locked phases~\cite{Alford_2007BV,Manuel_2007, Alford_2008BV,Bierkandt_2011,Mannarelli_2010}, being of particular interest the 2SC phase, which produce a similar bulk-viscous profile in the low-temperature regime as the normal phase \textcolor{black}{(see Fig. 5 in Ref.~\cite{Schmitt_2018})}.\\

To conclude, tidal dissipation effects during binary inspiral offer a unique possibility to probe transport properties of dense matter where conservative tidal effects (adiabatic and dynamical) only probe the equilibrium EoS of dense matter. For low mass compact binary systems ($\sim 1.4M_{\odot}$), a detection of tidal dissipation in gravitational waves can be a `smoking gun' signature for the presence of strange quark stars since there is no known nuclear matter source of high dissipation in that regime. Given their astrophysical significance and considering the increased sensitivity of next generation GW detectors, accurate BNS waveforms should be developed in near future incorporating tidal dissipation effects.    

\section*{Acknowledgments}
\textcolor{black}{D.C. would like to thank Prof. J. Schaffner-Bielich for pointing to her the article by Hernandez~et~al (2024) which led to this work.} The authors thank Swarnim Shirke, Nils Andersson and Alexander Haber for helpful discussions. SG acknowledges support from STFC via grant no. ST/Y00082X/1. We also acknowledge support from the program Unidad de Excelencia María de Maeztu CEX2020-001058-M, from the project PID2022-139427NB-I00 financed by the Spanish MCIN/AEI/10.13039/501100011033/FEDER, UE (FSE+),  as well as from the Generalitat de Catalunya under contract 2021 SGR 171, and by the CRC-TR 211 'Strong-interaction matter under extreme conditions'- project Nr. 315477589 - TRR 211. B.K.P acknowledges the use of IUCAA Pegasus cluster for the computational work and also the support of the CNRS-IN2P3 MAC masterproject, and the project RELANSE ANR-23-CE31-0027-01 of the French National Research Agency (ANR). This work makes use of \texttt{NumPy} \cite{vanderWalt:2011bqk}, \texttt{SciPy} \cite{Virtanen:2019joe}, \texttt{Matplotlib} \cite{Hunter:2007}, \texttt{jupyter} \cite{jupyter}, and \texttt{Corner} software packages.

\newpage
\bibliography{qm_tidal}

\newpage
\appendix
\section{Solving for the $f$-mode frequency within Relativistic Cowling Approximation}\label{app:f_mode}
Non-radial oscillation modes have been studied for many decades. The procedure for analysing these modes in the non-relativistic framework was \textcolor{black}{first discussed by Cowling~\cite{1942Obs....64..224C}, while in the context of general relativity, Thorne and Campollataro~\cite{Thorne1967} explored this further}. In general relativity, it is necessary to include the metric perturbation to solve the perturbed fluid equations. However, within the Cowling approximation, the metric perturbations can be neglected. It was demonstrated that the oscillation frequencies of the $f$-mode obtained using the Cowling approximation and those calculated using the complete linearized equations of general relativity differ by less than 20\%. As outlined in the main manuscript, we will adopt the Cowling approximation, where the spacetime metric for a spherically symmetric background is given by:
 \begin{equation}
     ds^2=e^{2\nu (r)}dt^2+e^{2\Lambda (r)}dr^2+r^2 d\theta^2+r^2\sin{\theta}^2 d\phi^2.\label{eqn:metric}
 \end{equation}
In order to find mode frequencies one has to solve the following differential equations:
\\
\sg{
\begin{eqnarray}
    \frac{d W(r)}{dr}&=&\frac{d \epsilon}{dp}\left[\omega^2r^2e^{\Lambda (r)-2\nu (r)}V (r)+\frac{d \nu(r)}{dr} W (r)\right] \nonumber \\
    &-& l(l+1)e^{\Lambda (r)}V (r), \nonumber \\
    \frac{d V(r)}{dr} &=& 2\frac{d\nu (r)}{dr} V (r)-\frac{1}{r^2}e^{\Lambda (r)}W (r), 
    \label{eqn:perteq}
\end{eqnarray}
}
\textcolor{black}{where
\begin{equation}\label{eq.gravphase}
    \frac{d \nu(r)}{dr}=-\frac{1}{\rho(r)+p(r)}\frac{dp}{dr},
\end{equation}
with $p$ being the pressure and $\rho$ the energy density as stated above.}

The functions $V (r)$ and $W (r)$ along with frequency $\omega$, characterize the Lagrange displacement vector $\mathcal{\xi}^i$ associated to a perturbed fluid
\begin{equation}
    \xi^i=\left ( \frac{W (r)}{e^{\Lambda (r)}},-V (r)\partial_{\theta},-\frac{V (r)}{ \sin^{2}{\theta}} \partial _{\phi}\right) \frac{Y_{lm}(\theta,\phi)}{r^2} ,
\end{equation} 
where $Y_{lm} (\theta,\phi)$ is the  $lm$-spherical harmonic. 

A solution of Eq.~(\ref{eqn:perteq}) with the fixed background metric in Eq.~(\ref{eqn:metric}) near the origin will behave as follows:
\begin{equation}
    W (r)=Ar^{l+1}, \> V (r)=-\frac{A}{l} r^l.
\end{equation}
The vanishing  perturbed Lagrangian pressure at the surface will provide another constraint to be included while solving Eq.~\ref{eqn:perteq}, which is given by
\begin{equation} 
    \omega^2e^{\Lambda (R)-2\nu (R)}V (R)+\frac{1}{R^2}\frac{d\nu (r)}{dr}\Big|_{r=R}W (R)=0.
\label{eqn:bc}
\end{equation}
Eqs.~(\ref{eqn:perteq}) are eigenvalue equations. Among the solutions, those $\omega^2$ that satisfy the boundary condition given by Eq.~(\ref{eqn:bc}) are the eigenfrequencies of the star. The eigenfrequency $\omega$, for which there is no radial node in the eigenfunction represents the $f$-mode frequency of the star. 

\section{Neutrino emissivity of quark matter}
\label{app:neutrino_emissivity}

In unpaired three-flavor quark matter the emission of neutrinos is given by the following semileptonic processes:
\begin{eqnarray}
\label{EW-semi-processes}
&& u+e^- \rightarrow d+\nu_e \ , \nonumber \\
&& d \rightarrow u+e^-+\bar{\nu}_e \ , \nonumber\\
&& u+e^- \rightarrow s+\nu_e \ ,\nonumber \\
&& s \rightarrow u+e^-+\bar{\nu}_e \ .
\end{eqnarray}
However, the main contribution is associated to the electroweak processes which involve the down quark. The different dependence on the Cabbibo angle suppresses the contribution of the processes that involve the strange quark (proportional to $\sin^2{\Theta_C}$). Neglecting this contribution, the neutrino emissivity can be computed as in Ref.~\cite{Iwamoto_1980,PhysRevD.70.114037} assuming the light quarks and electrons to be massless so that
\begin{equation}
    \mathbf{\epsilon}=\frac{457}{630}G_{F}^2\cos^2{\Theta_C}\alpha_s \mu_d \mu_u\mu_e T^6,
\end{equation}
where non-Fermi liquid effects are considered up to $\mathcal{O}(\alpha_s)$.

\textcolor{black}{The cooling in a strange quark star due to neutrino emissivity and neglecting surface emission can be studied assuming the medium to be isothermal given that thermal equilibration in degenerate matter is fast~\cite{PhysRevD.70.114037,PhysRevD.48.2916,Schmitt_2018}. Thus, the thermal evolution equation is given by
\begin{equation}
    C_V \frac{d\tilde{T}}{dt}=-L,
\end{equation}
where the average neutrino luminosity $L$ is computed over the entire stellar volume~\cite{Ofengeim:2017xxr}
\begin{equation}
    L(\tilde{T})=\int_{0}^{R}\epsilon(\rho,\tilde{T})e^{2\nu} \frac{4\pi r^2 dr}{\sqrt{1-2m/r}}.
\end{equation}
In the isothermal approximation, the average neutrino luminosity can be written as the following equation
\begin{equation}
    L=\upsilon \tilde{T}^6,
\end{equation}
where $\upsilon$ depends on stellar volume and the equal-mass configuration.}
\textcolor{black}{Using the simplified equations for the average heat capacity (see Eq.\eqref{eq.spheatcap}) and neutrino luminosity in the isothermal approximation, we obtain an expression for the final redshifted temperature, $\tilde{T}_f$ 
\begin{equation}\label{eq.tempneutrinos}
    \frac{1}{\tilde{T}_f^4}=\frac{1}{\tilde{T}_0^4}+\frac{4\upsilon t}{\sigma}.
\end{equation}
Here $t$ is the time that takes the system to reach a final temperature $\tilde{T}_f$ from an initial temperature $\tilde{T}_0$.}

\begin{figure}[H]
\includegraphics[width=0.483\textwidth]{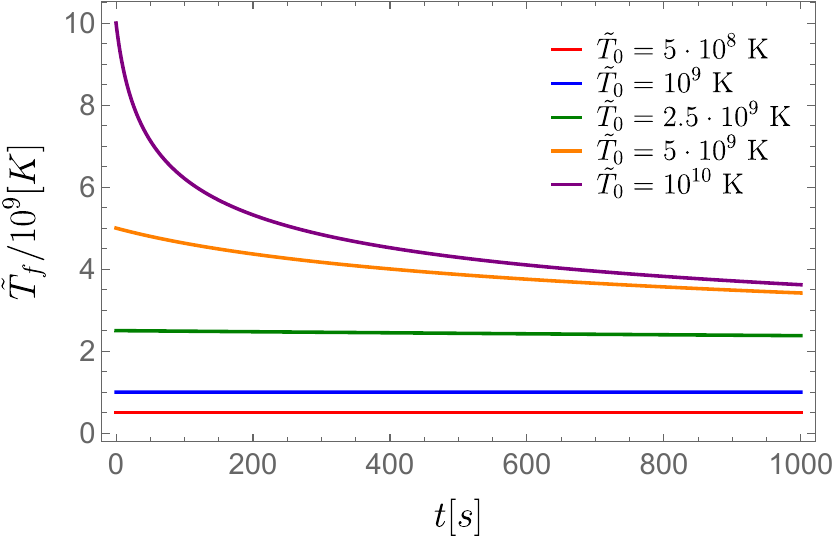}
\captionsetup{justification=centerlast}
\centering
\caption{\textcolor{black}{Thermal evolution of strange stars. The final redshifted temperature is normalized to $10^9$K and is shown as a function of the time (in seconds) for five initial temperatures \textcolor{black}{($\tilde{T}_0=$$5\cdot 10^8,\,10^9,\,2.5\cdot10^9,\,5\cdot10^9,\,10^{10}$ K)}, with $\alpha_s\approx 0.31$ and $m_s= 150$ MeV.}}
\label{fig.thermalevolneutrinos}
\end{figure}

In Fig.~\ref{fig.thermalevolneutrinos} we show the thermal evolution of strange stars employing Eq.~\eqref{eq.tempneutrinos}, {\color{black} starting from five different initial temperatures.} \textcolor{black}{We consider an equal-mass binary of $1.8M_{\odot}$ and compute $\upsilon$ considering the density profiles (and the chemical potential variations with density) obtained by solving the TOV equations with the EoS of the non-ideal bag model (we use an approximated value of $\alpha_s$, so that $a_4=1-2\alpha_s/\pi$\cite{Fraga_2001}). In this case we get $\upsilon\approx 1.86\times 10^{-12}$ erg/(s$\cdot$K$^6$)
and $\sigma\approx 1.30\times 10^{30 }$ erg/K$^2$. As can be seen, the cooling due to neutrino emission is negligible for temperatures below $2.5\times 10^9$ K in the inspiral timescale of $100$ s.} Therefore, its contribution is small for the study of the tidal heating of strange quark stars.

\end{document}